\documentclass[12pt]{article}

\usepackage{amsfonts}
\usepackage{amsmath}
\usepackage{epsfig}
\usepackage{latexsym}
\usepackage{graphicx}
\usepackage{amssymb}
\usepackage{color}
\numberwithin{equation}{section}
\usepackage{url}
\allowdisplaybreaks

\def\spa#1{\phantom{\fbox{\rule[-#1cm]{0cm}{0cm}}}}

\def\be{\begin{equation}}
\def\ee{\end{equation}}
\def\bea{\begin{eqnarray}}
\def\eea{\end{eqnarray}}
\def\bequ{\begin{equation}}
\def\eequ{\end{equation}}

\def\del{\partial}

\renewcommand{\thefootnote}{\fnsymbol{footnote}}

\newcommand{\eq} {equation}
\newcommand{\eqa} {eqnarray}
\newcommand{\NN} {\mbox {$\nonumber$}}

\begin{document}

\hfuzz=100pt
\title{Identification of Bulk coupling constant\\
in Higher Spin/ABJ correspondence}
\date{}
\author{Masazumi Honda\footnote{masazumihondaAThri.res.in}
  \spa{0.5} \\
\\
{\small{\it Harish-Chandra Research Institute,}}
\\ {\small{\it Chhatnag Road, Jhusi, Allahabad 211019, India}} \\
}
\date{\small{June 2015}}

\maketitle
\thispagestyle{empty}
\centerline{}

\begin{abstract}
We study the conjectured duality
between the $\mathcal{N}=6$ Vasiliev higher spin theory on $AdS_4$ and 
3d $\mathcal{N}=6$ superconformal Chern-Simons matter theory known as the ABJ theory.
We discuss how the parameters in the ABJ theory should be related to the bulk coupling constant in the Vasiliev theory.
For this purpose,
we compute two-point function of stress tensor in the ABJ theory by using supersymmmetry localization.
Our result justifies the proposal by arXiv:1504.00365 and
determine the unknown coefficient in the previous work.
\end{abstract}
\vfill
\noindent HRI/ST/1506 

\renewcommand{\thefootnote}{\arabic{footnote}}
\setcounter{footnote}{0}

\newpage
\setcounter{page}{1}
\section{Introduction}
It has been expected that
string theory at extremely high energy possesses huge symmetry generated by infinite massless higher spin fields \cite{Gross:1988ue}.
While the Vasiliev theories \cite{Vasiliev:2003ev} are known as consistent interacting theories of massless higher spin gauge fields,
it is still unclear 
how the Vasiliev theories are related to tensionless limit of string theory.
Nevertheless the Vasiliev theories have recently provided great interests in the context of AdS/CFT correspondence \cite{Maldacena:1997re}
as initiated in \cite{Klebanov:2002ja}.
Interestingly, in such higher spin version of AdS/CFT correspondences,
their dual CFT sides sometimes have clear origins from string theory\footnote{
See e.g. \cite{Gaberdiel:2013vva,Creutzig:2014ula,Gaberdiel:2014cha,Gaberdiel:2015mra}
in the context of higher spin $AdS_3$/$CFT_2$ correspondence.
}.
It would give some new insights to a relation between the Vasiliev theory and string theory
if we study this type of correspondence.

A good laboratory for this purpose is provided by
the ABJ theory \cite{Aharony:2008gk,Aharony:2008ug},
which is 3d $\mathcal{N}=6$ superconformal Chern-Simons matter theory 
with the gauge group $U(N)_k \times U(N+M)_{-k}$ and the Chern-Simons level $k$.
The ABJ theory is expected
as the low-energy effective theory of $N$ M2-branes probing $\mathbf{C}^4 /\mathbb{Z}_k$
and $M$ fractional M2-branes sitting at the singularity.
In usual story of the AdS/CFT,
the ABJ theory is expected to be dual to M-theory on $AdS_4 \times S^7 /\mathbf{Z}_k$ 
with the 3-form $C_3 \propto M/k $ and
type IIA superstring on $AdS_4 \times\mathbb{CP}^3$ 
with the NS-NS 2-form $B_2 \propto M/k $. 
Here we consider apparently different type of the AdS/CFT correspondence.
It was recently proposed that 
the ABJ theory is well described by 
$\mathcal{N}=6$ Vasiliev theory on $AdS_4$ in the limit \cite{Giombi:2011kc,Chang:2012kt}
\begin{\eq}
M\gg 1 ,\quad k\gg 1 ,\quad N={\rm fixed},\quad  t=\frac{M}{k} ={\rm fixed} .
\label{eq:HSlimit}
\end{\eq}
After a while,
the authors in \cite{Hirano:2015yha} have precisely tested 
this proposal for partition function on $S^3$.
On the boundary side,
they have developed
the systematic $1/M$ expansion of the ABJ partition function using the previous results \cite{Awata:2012jb,Honda:2013pea,Honda:2014npa}.
On the bulk side,
they have computed the one-loop free energy of the Vasiliev theory
by using the technique in \cite{Giombi:2013fka,Giombi:2014iua}.
For comparing the both results,
they proposed that 
the bulk coupling constant $G_{\rm HS}$, namely the Newton constant in the Vasiliev theory,
is related to the parameters in the ABJ theory by\footnote{
We are taking unit AdS radius.
} \cite{Hirano:2015yha} 
\begin{\eq}
G_{\rm HS} =\frac{\gamma}{M} \frac{\pi t}{\sin{(\pi t)}} ,
\label{eq:newton}
\end{\eq}
with the unknown coefficient $\gamma$.

In this paper we justify this identification \eqref{eq:newton} 
and determine the value of the unknown coefficient $\gamma$.
Namely, we discuss how the parameters in the ABJ theory should be related to the bulk coupling constant in the Vasiliev theory.
This problem is essentially equivalent to find the parameter
``$\tilde{N}$" in Maldacena-Zhiboedov \cite{Maldacena:2011jn,Maldacena:2012sf},
which is the natural expansion parameter in 3d theories 
with (slightly broken) higher spin symmetries.
For this purpose,
we compute two-point function of stress tensor in the ABJ theory by using supersymmmetry localization \cite{Pestun:2007rz}.
Finally, we will show
\begin{\eq}
G_{\rm HS}  = \frac{2t}{M\sin{(\pi t)}} ,\quad  \gamma =\frac{2}{\pi} ,
\label{eq:main}
\end{\eq}
in the canonical normalizations on the both sides.

This paper is organized as follows.
In sec.~\ref{sec:localization},
we explain how to compute 
the stress tensor two-point function by using the localization.
In sec.~\ref{sec:derivation},
we compute the two-point function in the higher spin limit and
derive our main result \eqref{eq:main}.
Section \ref{sec:conclusion} is devoted to conclusion and discussions.

\section{Two point function of stress tensor from localization}
\label{sec:localization}
In this section,
we discuss how one can compute 
the stress tensor two-point function in the ABJ theory by using the localization.
In 3d CFT,
two point function of canonically normalized stress tensor in flat space at separate points
takes the form \cite{Osborn:1993cr}
\begin{\eq}
\langle T_{\mu\nu}(x) T_{\rho\sigma}(0) \rangle
=\frac{c_T}{64} (P_{\mu\rho}P_{\nu\sigma} +P_{\nu\rho}P_{\mu\sigma} -P_{\mu\nu}P_{\rho\sigma}) \frac{1}{16\pi^2 x^2},
\end{\eq}
where $P_{\mu\nu}=\delta_{\mu\nu}\del^2 -\del_\mu \del_\nu$.
We normalize $c_T$
such that one free real scalar and Majorana fermion contribute to $c_T$ by $c_T =1$.
Here we would like to compute $c_T$ of the ABJ theory 
in the higher spin limit \eqref{eq:HSlimit}.
This limit is equivalent to treat 
one of the 't Hooft couplings as the small parameter 
but to keep the other 't Hooft coupling finite.
Hence it is nice 
if we can compute $c_T$ by using some non-perturbative methods.

Fortunately there are two ways to compute $c_T$ 
by using the SUSY localization \cite{Pestun:2007rz}.
One way \cite{Closset:2012ru} is
to get $c_T$ from partition function of the ABJ theory on squashed sphere\footnote{
Strictly speaking, this squashed sphere should be so-called bi-axially squashed sphere 
in original setup.
Although there are many other choices of squashed $S^3$,
we have the same partition function \cite{Closset:2013vra} 
as long as we consider one-parameter deformation of the round sphere with keeping SUSY.
} \cite{Hama:2011ea,Imamura:2011wg}.
This method has been applied to various examples in \cite{Nishioka:2013gza}.
The other way is to compute two-point function of flavor symmetry current by the localization \cite{Closset:2012vg}
and then find $c_T$ from its coefficient.
Here we take the latter approach.

In 3d $\mathcal{N}=2$ language,
the ABJ theory consists of vector multiplet, two bi-fundamental chiral multiplets $(A_1 ,A_2)$ and
two anti-bi-fundamental chiral multiplets $(B_1 ,B_2 )$ 
with the superpotential \cite{Aharony:2008gk,Aharony:2008ug}
\begin{\eq}
W\sim \epsilon^{\alpha\beta}\epsilon^{\gamma\delta} {\rm Tr}[ A_\alpha B_\gamma A_\beta B_\delta ] ,
\end{\eq}
where $\alpha ,\beta ,\gamma ,\delta =1,2$.
Let us consider a particular $U(1)_f$ flavor symmetry 
summarized in table \ref{tab:flavor}.
Then, the two-point function of the flavor symmetry current $j_\mu$ 
is fixed by the 3d conformal symmetry as\footnote{
We have neglected the contact term.}:
\begin{\eq}
\langle j_a^\mu (x) j_b^\nu (0) \rangle
=\frac{\tau_f}{16\pi^2} (\delta^{\mu\nu}\del^2 -\del^\mu \del^\nu ) \frac{1}{x^2} ,
\end{\eq}
up to the unknown coefficient\footnote{
$\tau_f$ is the same as $\tau_{22}$ in the notation of \cite{Chester:2014fya}.
} $\tau_f$.
The coefficient $\tau_f$ is proportional to $c_T$ and
its proportional coefficient for the ABJ theory has been fixed as \cite{Chester:2014fya}
\begin{\eq}
c_T = 4\tau_f .
\end{\eq}

\begin{table}[tbp]
\begin{center}
  \begin{tabular}{|c|c  c c c |}
  \hline              & $A_1$       &$A_2$   & $B_1$ &  $B_2$ \\
\hline $U(1)_f$   & +1            & -1        & +1      &   -1      \\  \hline
  \end{tabular}
\end{center}
\caption{Charge assignments of $U(1)_f$ flavor symmetry.}
\label{tab:flavor}
\end{table}

We can compute $\tau_f$ by using the localization in the following way.
First we introduce supersymmetric flavor mass $m$ of the $U(1)_f$ symmetry 
by weakly gauging this symmetry and turning on its fixed Coulomb branch\footnote{
This corresponds to just fix the adjoint scalar in the $U(1)_f$ vector multiplet to the constant. 
}.
If we denote the partition function of the mass-deformed ABJ theory on $S^3$ by $Z(m)$,
then the partition function $Z(m)$ generates $\tau_f$ by the relation\footnote{
We have rescaled the mass as $m\rightarrow m/(2\pi)$.
} \cite{Closset:2012vg}
\begin{\eq}
\tau_f = -8 \left. {\rm Re}\frac{1}{Z(0)}\frac{\del^2 Z(m)}{\del m^2} \right|_{m=0} .
\end{\eq}
Since the mass-deformed ABJ theory still has at least $\mathcal{N}=2$ SUSY,
we can compute $Z(m)$ by the localization\footnote{
This matrix model was analyzed in \cite{Anderson:2014hxa,Anderson:2015ioa,Drukker:2015awa} in different contexts for $N_1 =N_2$ (ABJM case).
} \cite{Kapustin:2009kz,Jafferis:2010un,Hama:2010av}:
\begin{\eqa}
Z(m)
&=& \frac{1}{N_1 ! N_2 !} \int \frac{d^{N_1}\mu}{(2\pi )^{N_1} } \frac{d^{N_2}\nu}{ (2\pi )^{N_2}}
e^{\frac{ik}{4\pi} \sum_{j=1}^{N_1} \mu_j^2  -\frac{ik}{4\pi} \sum_{a=1}^{N_2} \nu_a^2 } \NN\\
&&\times \frac{ \prod_{1\leq i \neq j \leq N_1 } 2\sinh{\frac{\mu_i -\mu_j }{2}}   
\prod_{1\leq a\neq b \leq N_2}  2\sinh{\frac{\nu_a -\nu_b}{2}}   }
 {\prod_{j=1}^{N_1}\prod_{a=1}^{N_2} 2\cosh{\frac{\mu_j -\nu_a +m}{2}}  \cdot 2\cosh{\frac{\mu_j -\nu_a -m}{2}}  } ,
\end{\eqa}
where $N_1 =N$ and $N_2 =N+M$.
In this way, we can compute $c_T$ by using the localization.
In next section we compute $Z(m)$ and find $c_T$ in the higher spin limit.

\section{Derivation}
\label{sec:derivation}
We would like to consider the higher spin limit
\[
M\gg 1 ,\quad k\gg 1 ,\quad N={\rm fixed},\quad  t=\frac{M}{k} ={\rm fixed} .
\]
However, instead we first consider the slightly different limit:
\begin{\eq}
N_2 \gg 1 , k\gg 1 ,\quad \lambda_1 =\frac{N_1}{k}= \frac{t N}{M} \ll 1 ,\quad 
\lambda_2 =\frac{N_2}{k}=\left( 1+\frac{N}{M} \right) t ={\rm fixed} ,
\label{eq:conifold}
\end{\eq}
which corresponds to the $1/N_2$ expansion.
Then we will perform $1/M$ expansion of the result in the limit \eqref{eq:conifold} 
and extract the higher spin limit.
For this purpose, it is convenient to rewrite 
the mass-deformed partition function as
\begin{\eqa}
Z(m)
&=& \frac{1}{ N_1 ! } \int \frac{d^{N_1}\mu}{(2\pi )^{N_1} }
e^{\frac{ik}{4\pi} \sum_{j=1}^{N_1} \mu_j^2 } 
  \prod_{ i \neq j } (\mu_i -\mu_j ) \left\langle  e^{V(\mu ,\nu )} \right\rangle_{N_2} ,
\end{\eqa}
where
\begin{\eq}
V(\mu ,\nu )
= \sum_{i\neq j} \log{\frac{2\sinh{\frac{  \mu_i -\mu_j }{2}}}{\mu_i -\mu_j}} 
-\sum_{j,a}\Biggl[ \log{\left( 2\cosh{\frac{\mu_j -\nu_a +m}{2}} \right)} +\log{\left( 2\cosh{\frac{\mu_j -\nu_a -m}{2}} \right)} \Biggr] .
\end{\eq}
The symbol $\langle\mathcal{O}\rangle_{N_2 }$ denotes
the unnormalized VEV over the $U(N_2 )$ part: 
\begin{\eq}
\langle\mathcal{O}\rangle_{N_2}
=\frac{1}{N_2 !} \int  \frac{d^{N_2}\nu}{ (2\pi )^{N_2}}
\mathcal{O}e^{-\frac{1}{2g_s} \sum_a \nu_a^2 }
 \prod_{ a \neq b } \Biggl[  2\sinh{\frac{\nu_a -\nu_b }{2}}  \Biggr] ,\quad
{\rm with}\ g_s =-\frac{2\pi i}{k} ,
\end{\eq}
which is the same as the VEV in the $U(N_2 )_{-k}$ CS matrix model 
on $S^3$ (without level shift).
When the first 't Hooft coupling $\lambda_1$ is small,
the integral over $\mu$ is dominated by $\mu\simeq 0$ and
we can approximate $V(\mu ,\nu )$ by small $\mu$ expansion as usual:
\begin{\eqa}
V(\mu, \nu )
&=& -N_1 \sum_{a}\Biggl[ \log{\left( 2\cosh{\frac{ \nu_a +m}{2}} \right)}
+\log{\left( 2\cosh{\frac{ \nu_a -m}{2}} \right)} \Biggr] \NN\\
&& +\sum_{j}\mu_j \sum_a \tanh{\frac{ \nu_a +m}{2}} 
+\sum_j \mu_j \sum_a \tanh{\frac{  \nu_a -m}{2}} +\mathcal{O}(\mu^2 ) \NN\\
&=& -N_1 \sum_{a}\Biggl[ \log{\left( 1+e^{  \nu_a +m}  \right)}
+\log{\left( 1+e^{  \nu_a -m}  \right)} -\nu_a  \Biggr] \NN\\
&& +\sum_{j}\mu_j \sum_a \tanh{\frac{ \nu_a +m}{2}} 
+\sum_j \mu_j \sum_a \tanh{\frac{  \nu_a -m}{2}} +\mathcal{O}(\mu^2 ) .
\end{\eqa}
Since the integrand is symmetric under $\mu \rightarrow -\mu$ and $\nu \rightarrow -\nu$, 
we find
\begin{\eqa}
Z(m)
&=& \frac{1}{ N_1 ! } \int \frac{d^{N_1}\mu}{(2\pi )^{N_1} } 
e^{\frac{ik}{4\pi} \sum_{j=1}^{N_1} \mu_j^2 } 
  \prod_{ i \neq j } (\mu_i -\mu_j )  \NN\\
&&\times  \left\langle  \exp\Bigl[ -N_1 \sum_a \log (1+e^{ \nu_a +m})(1+e^{ \nu_a -m}) \Bigr] 
 +\mathcal{O}(\mu^2 ) \right\rangle_{N_2} .
\end{\eqa}
This can be computed by using the technique in \cite{Drukker:2010nc} 
used for conifold expansion of the ABJ(M) theory.
Let us introduce the quantity
\begin{\eq}
g(Y) 
=-g_s  \Bigl\langle \sum_a  \log{( 1-Ye^{\nu_a} )} \Bigr\rangle_{N_2 ,{\rm planar}} ,
\end{\eq}
where $\langle \cdots \rangle_{N_2 ,{\rm planar}}$ denotes
the VEV in the planar limit for the $U(N_2 )$ gauge group.
Then, in the higher spin limit \eqref{eq:HSlimit}, we find
\begin{\eq}
\langle e^{V (\mu ,\nu)} \rangle_{N_2 }
\simeq    \exp\Biggl[ \frac{N_1 }{g_s} \left(  g(-e^m) +g(-e^{-m}) \right) \Biggr]  .
\end{\eq}
Fortunately the quantity $g(Y)$ has been computed in \cite{Drukker:2010nc} for arbitrary $Y$ as
\begin{\eqa}
g(Y) 
&=& \frac{\pi^2}{6} -\frac{1}{2} \left( \log{h(Y)} \right)^2 
+\log{h(Y)} \left( \log{(1-e^{-t_2}h(Y))} -\log{(1-h(Y))} \right) \NN\\
&& -{\rm Li}_2 (h(Y)) +{\rm Li}_2 (e^{-t_2}h(Y)) -{\rm Li}_2 (e^{-t_2}) ,
\end{\eqa}
where
\begin{\eq}
t_2 =-\frac{2\pi i N_2}{k},\quad 
h(Y) = \frac{1}{2} \left[ 1+Y +\sqrt{(1+Y)^2 -4e^{t_2}Y} \right] .
\end{\eq}
Thanks to this result, one can show
\begin{\eq}
\left. \frac{\del^2}{\del m^2}\langle e^{V(\mu ,\nu )} \rangle_{N_2 } \right|_{m=0}
\simeq  -\frac{N_1}{g_s} (1-e^{-\frac{t_2}{2}})   e^{ \frac{2N_1}{g_s}   g(-1)}  
= -\frac{N M}{\pi t} e^{\frac{\pi it}{2}} \sin{\frac{\pi t}{2}}  e^{ \frac{2N_1}{g_s}   g(-1)} 
+\mathcal{O}(1)   .
\label{eq:VEV}
\end{\eq}
Noting that $Z(0)$ in the limit \eqref{eq:conifold} is given by
\begin{\eq}
Z(0)
\simeq \frac{1}{ N_1 ! } \int \frac{d^{N_1}\mu}{(2\pi )^{N_1} } 
e^{ \frac{2N_1 }{g_s} g(-1) } e^{\frac{ik}{4\pi} \sum_{j=1}^{N_1} \mu_j^2 } 
     \prod_{ i \neq j } (\mu_i -\mu_j ) ,
\end{\eq}
we finally obtain
\begin{\eq}
c_T 
= -32 {\rm Re}\left( -\frac{N M}{\pi t} e^{\frac{\pi it}{2}} \sin{\frac{\pi t}{2}}   \right) 
=  \frac{16N M\sin{\pi t}}{\pi t} .
\end{\eq}
As a simple consistency check, let us consider the $t\rightarrow 0$ limit.
Then, since the ABJ theory has $8N(N+M)$ real scalars and $8N(N+M)$ Majorana fermions,
$c_T$ should be $16N(N+M)=16NM+\mathcal{O}(1)$.
Our result is actually consistent with this result.
As a conclusion, if we take the canonical normalization\footnote{
More precisely we suppose to take the normalization such that
if we knew quadratic ``actions" for the spin-2 field fluctuations in the dual Vasiliev theory, 
then the spin-2 field ``actions" are the same as the one for the canonically normalized Einstein gravity in $AdS_4$
with identifying $G_{\rm HS}$ with the 4d Newton constant.
} for spin-2 fields (see e.g. \cite{Buchel:2009sk}) and
note that the stress tensor corresponds to $U(N)$ singlet \cite{Chang:2012kt} on the gravity side ,
then the bulk coupling constant $G_{\rm HS}$ should be given by
\begin{\eq}
G_{\rm HS} = \frac{32 N}{\pi c_T}   = \frac{2t}{M\sin{(\pi t)}} .
\end{\eq}
This determines the unknown coefficient $\gamma$ in \eqref{eq:newton} of the previous study \cite{Hirano:2015yha}
as $\gamma =2/\pi $. 

\section{Conclusion and discussions}
\label{sec:conclusion}
We have focused on the conjectured duality
between the $\mathcal{N}=6$ Vasiliev higher spin theory on $AdS_4$ and 
the ABJ theory \cite{Giombi:2011kc,Chang:2012kt}.
We have discussed how the parameters in the ABJ theory should be related 
to the bulk coupling constant $G_{\rm HS}$ in the Vasiliev theory.
To achieve this,
we have computed the two-point function of the stress tensor in the ABJ theory 
by using the supersymmmetry localization.
As a result, we have justified the identification \eqref{eq:newton} proposed in \cite{Hirano:2015yha}
and determined the value of the unknown coefficient $\gamma$ as \eqref{eq:main}.
Our result on $c_T$ is similar to the previous results \cite{Aharony:2012nh,GurAri:2012is}
on non-supersymmetric $U(M)_k$ Chern-Simons theory with fundamental matters:
\begin{\eq}
c_{T,\rm fund.} = \frac{2M\sin{(\pi t)}}{\pi t} ,
\end{\eq}
where $t=M/k_{\rm eff}$ with the effective CS level $k_{\rm eff}$.
It would be interesting to understand why the factor $\sin{(\pi t)}/(\pi t)$ so universally appears.

Besides the higher spin limit,
it is also illuminating to study $c_T$ 
in the context of the usual AdS/CFT correspondence
between the ABJ(M) theory and M-theory or type IIA superstring.
It is known that
the partition function of the mass-deformed ABJM theory on $S^3$
is described by an ideal Fermi gas \cite{Drukker:2015awa,Marino:2011eh}.
Probably we can show that 
the ABJ case ($M\neq 0$) also has an ideal Fermi gas picture 
by using the technique in \cite{Honda:2013pea}.
Then we should be able to study
non-perturbative corrections \cite{Cagnazzo:2009zh,Drukker:2011zy} in M-theory 
to the stress tensor two-point function
as in the partition function \cite{Hatsuda:2012dt,Calvo:2012du,Hatsuda:2013gj,Hatsuda:2013oxa,Matsumoto:2013nya,Honda:2014npa}
and supersymmetric Wilson loops \cite{Grassi:2013qva,Hatsuda:2013yua}.
We expect that
this approach can also precisely test the conjecture $c_T \geq 16$ for 3d $\mathcal{N}=8$ SCFT's
from conformal bootstrap \cite{Chester:2014fya}.

We close by a comment to contact term in the flavor current two-point function
discussed in \cite{Closset:2012vg,Closset:2012vp}:
\begin{\eq}
\langle j_a^\mu (x) j_b^\nu (0) \rangle
=\frac{\tau_f}{16\pi^2} (\delta^{\mu\nu}\del^2 -\del^\mu \del^\nu ) \frac{1}{x^2} 
+\frac{i\kappa_f}{2\pi}\epsilon^{\mu\nu\rho} \del_\rho \delta^{(3)}(x) .
\end{\eq}
We can compute the coefficient $\kappa_f$ by \cite{Closset:2012vg,Closset:2012vp}
\begin{\eq}
\kappa_f = 2\pi \left. {\rm Im}\frac{1}{Z(0)}\frac{\del^2 Z(m)}{\del m^2} \right|_{m=0} .
\end{\eq}
Looking at \eqref{eq:VEV}, we immediately find $\kappa_f$ in the higher spin limit as
\begin{\eq}
\kappa_f = -\frac{2M}{t} \sin^2{\frac{\pi t}{2}} .
\end{\eq}
It is attractive if we find physical interpretations of this formula from the gravity side.

\subsection*{Acknowledgment}
This work is motivated by our previous collaboration \cite{Hirano:2015yha}
with Shinji Hirano, Kazumi Okuyama and Masaki Shigemori.
We would like to thank Rajesh Gopakumar, Shinji Hirano, Kazumi Okuyama and Yuki Yokokura for useful discussions.
We are grateful to ICTS for warm hospitality, where a part of this work was done.

\providecommand{\href}[2]{#2}\begingroup\raggedright\endgroup

\end{document}